\documentclass[11pt]{article}

\usepackage[T2A]{fontenc}
\usepackage[cp866]{inputenc}
\usepackage{amsthm,amssymb}
\usepackage{graphicx}

\usepackage{graphics}

\makeatletter \@addtoreset{equation}{section} \makeatother

\newtheorem{theorem}{Theorem}
\newtheorem{lemma}{Lemma}

\newtheorem{definition}{Definition}

\def\fracd{\displaystyle\frac}
\def\sumd{\displaystyle\sum}
\def\limd{\displaystyle\lim}

\begin{document}
\baselineskip 23pt \vskip 0.5 cm

\title
{EIGENVALUE DISTRIBUTION OF
LARGE WEIGHTED BIPARTITE RANDOM GRAPHS}
\author{
$\;$V. Vengerovsky  $\!\!\:\!$, Institute for Low Temperature Physics,
Ukraine}
\date{}
\maketitle
\abstract{
We study eigenvalue distribution of the adjacency
matrix $A^{(N,p, \alpha)}$
of weighted random bipartite graphs $\Gamma= \Gamma_{N,p}$. We assume
that the graphs have
$N$ vertices, the ratio of parts is $\fracd{\alpha}{1-\alpha}$ and  the average number of edges attached to one vertex  is
$\alpha\cdot p$ or $(1-\alpha)\cdot p$. To each edge of the graph $e_{ij}$ we assign a weight given by a random variable
$a_{ij}$ with  all moments finite.

We consider the moments of
normalized eigenvalue counting measure $\sigma_{N,p, \alpha}$  of
$A^{(N,p, \alpha)}$. The weak convergence in probability  of normalized eigenvalue counting measures is proved.

\section{Introduction}

The spectral theory of graphs is an actively developing field of
mathematics involving a variety of methods and deep results (see the
monographs \cite{Ch,CDS,GR}).
Given a graph with $N$ vertices, one can associate with it many
different matrices, but the most studied are the adjacency
matrix and the Laplacian matrix of the graph.
Commonly, the set of $N$ eigenvalues of the adjacency matrix
is referred to as the spectrum of the graph.
In these studies, the dimension of the matrix $N$
is usually regarded as a fixed parameter.
The spectra of infinite graphs  is considered
in certain particular cases of graphs having  one or another
regular structure (see for example \cite{JMRR}).

Another large class of graphs, where the limiting transition
$N\to\infty$ provides a natural approximation is represented
by random graphs \cite{B,JLR}. In this branch,
geometrical and topological properties of graphs (e.g. number of connected components, size of maximal connected component) are  studied for
immense number of random graph ensembles.
One of the classes of the
prime reference is the {\it binomial random graph}
originating by P. Erd\H{o}s (see, e.g. \cite{JLR}).
Given a number $p_N\in (0,1)$, this family of graphs ${\mathbf G}(N,p_N)$
is  defined by taking as $\Omega$ the set of all graphs
on $N$ vertices with the probability
$$
P(G) = p_N^{e(G)} (1-p_N)^{{N \choose 2} - e(G)},
\eqno (1.1)
$$
where $e(G)$ is the number of edges of $G$. Most of the random graphs
studies are devoted to the cases where $p_N\to 0$ as $N\to\infty$.

Intersection of these two branches of the theory of graphs contains
 the spectral theory of random graphs that is still poorly explored.
  However, a number of powerful tools can be employed here, because
 the ensemble of random symmetric $N\times N$ adjacency matrices
  $A_N$ is a  particular
representative of the random matrix theory,
where the limiting transition $N\to\infty$
is intensively studied during half of century since the pioneering
works by E. Wigner \cite{W}.
Initiated by theoretical physics applications,
the spectral theory of random matrices
has revealed deep nontrivial links with many
fields of mathematics.

Spectral properties of random  matrices corresponding to (1.1)
were examined in the limit $N\to\infty$
both in numerical and theoretical physics studies
\cite{E1, E2, E3,MF:91,RB:88,RD:90}.
There are  two major asymptotic regimes:
$p_N \gg 1/N$ and $p_N = O(1/N)$ and corresponding models
can be called the {\it dilute random matrices} and {\it sparse random
matrices}, respectively.
The first studies of
spectral properties of sparse and dilute random matrices
in the physical
literature are related with the works  \cite{RB:88}, \cite{RD:90},
\cite{MF:91}, where equations for the limiting density of states of
sparse random matrices  were
derived. In  papers \cite{MF:91} and \cite{FM:96} a number
of important  results on the universality of the correlation
functions
  and the Anderson localization transition were obtained.
  But all these results were obtained with
   non rigorous  replica and
supersymmetry methods.

On mathematical level of rigour, the  eigenvalue distribution of dilute
random matrices was studied in \cite{KKPS}. It was shown that
  the normalized eigenvalue counting function of
$$
{1\over \sqrt{N p_N} } A_{N,p_N}
\eqno (1.2)
$$
converges in the limit $N, p_N \to \infty$ to the distribution of explicit form known as the
semicircle, or Wigner law
\cite{W}.
The moments of this distribution verify well-known recurrent relation
for the Catalan numbers and can be found explicitly.
Therefore one can say that the dilute random matrices
represent explicitly solvable model (see also \cite{RB:88,RD:90}).

In the series of papers \cite{BG1,BG2,B} and  in \cite{KSV},
the adjacency  matrix and the Laplace matrix of random graphs (1.1)
with $p_N = pN$ were studied.
It was shown that
  this sparse random matrix ensemble can also be viewed as the explicitly
solvable model.

In the present paper we consider a bipartite analogue of  large sparse random  graph. This article is a modification of one part of \cite{KSV} for this case.


\section{Main results}


    We can introduce the randomly weighted
adjacency matrix of random bipartite graphs.
Let $\Xi=\{a_{ij} ,\; i \!\leq\! j,\;i,j\! \in \!{\bf N}\}$ be
the  set of
jointly independent identically distributed (i.i.d.) random
variables determined
  on the same probability space and possessing the moments
\begin{equation}\label{const_all_mom}
  {\bf E}a^{k}_{ij}\!=\!X_{k}\!<\!\infty \qquad
  \forall \; i,j,k\in {\bf N},
\end{equation}
where ${\bf E}$ denotes the mathematical expectation corresponding to
$\Xi$.
  We set $a_{ji}\!=\! a_{ij}$ for $i\!\leq\! j \;$.

Given $ 0\!<p\!\leq \!N \;$, let us define the family
  $D^{(p)}_N\!=\!\{d^{(N,p)}_{ij},
\; i\!\leq\! j,\; i,j\in \overline{1,N}\}$ of jointly independent
random variables
\begin{equation}
d^{(N,p)}_{ij}\!=\! \left\{ \begin{array}{ll} 1,&
\textrm{with} \ \textrm{probability } \ p/N ,
\\0,& \textrm{with} \ \textrm{probability} \ 1-p/N ,\\ \end{array}
\right.
\end{equation}
We determine $d_{ji}= d_{ij}$ and assume that $ \Lambda^{(p)}_N$
is independent from $\Xi$.

 Let $\alpha \in (0,1)$, define $I_{\alpha,N}=\overline{1,[\alpha \cdot N]}$, where $[\cdot]$ is a floor function. Now one can consider the real symmetric $N\times N$ matrix
$A^{(N,p,\alpha)}(\omega)$:
\begin{equation}\label{dilute}
\left[A^{(N,p,\alpha)}\right]_{ij}\!=\!\left\{ \begin{array}{ll} a_{ij}\cdot d_{ij}^{(N,p)},&
\textrm{if} \  (i \in I_{\alpha,N} \wedge j\notin I_{\alpha,N} ) \vee (i \notin I_{\alpha,N} \wedge j\in I_{\alpha,N} )  ,
\\0,& \textrm{otherwise} \\ \end{array}
\right.
\end{equation}
that has $N$ real eigenvalues
$\lambda^{(N,p,\alpha)}_1\!\leq\!\lambda^{(N,p,\alpha)}_2 \!\leq\!\ \ldots
\!\leq\!\ \lambda^{(N,p,\alpha)}_N$.

The normalized eigenvalue counting function (or integrated density
of states (IDS)) of  $A^{(N,p,\alpha)}$ is determined
by the formula
$$
\sigma\left({\lambda; A^{(N,p,\alpha)}}\right)\!=\!\frac{\sharp
\{j:\lambda^{(N,p,\alpha)}_j\!<\!\lambda\}}{N}.
$$

\begin{theorem}\label{main_thm}
Under condition
\begin{equation}\label{Karleman_cond}
X_{2m}\le \left(C \cdot m\right)^{2m}, m\in \mathbb{N}
\end{equation}
the measure $\sigma\left({\lambda; A^{(N,p,\alpha)}}\right)$ weak converges in probability to nonrandom measure $\sigma_{p,\alpha}$
\begin{equation}
\sigma\left({\cdot \ ; A^{(N,p,\alpha)}}\right)\to \sigma_{p,\alpha},\ N \to\infty ,
\end{equation}
which can be uniquely determine  by the moments
\begin{equation}\label{5}
\int \lambda^s d \sigma_{p,\alpha}=\left\{ \begin{array}{ll}
m^{(p,\alpha)}_{k}\!=\!\sum^{k}_{i=0}\left( S^{(1)}(k,i)+ S^{(2)}(k,i)\right),&
\textrm{if} \ s=2k ,
\\0,& \textrm{if } \ s=2k-1 ,\\ \end{array} \right.,
\end{equation}
where numbers $S^{(1)}(k,i)$,  $S^{(2)}(k,i)$ can be found from the following system of recurrent relations
\begin{equation}\label{ms_r}
  \mathrm{S}^{(1)}(l,r)=p\cdot \sum_{f=1}^{r} {r-1\choose f-1} \cdot
   X_{2f}
         \cdot \sum_{u=0}^{l-r} \mathrm{S}^{(1)}(l-u-f,r-f)
       \cdot \sum_{v=0}^{u} {f+v-1\choose f-1} \cdot
       \mathrm{S}^{(2)}(u,v)
\end{equation}
\begin{equation}\label{ms_r2}
   \mathrm{S}^{(2)}(l,r)=p\cdot \sum_{f=1}^{r} {r-1\choose f-1} \cdot
   X_{2f}
         \cdot \sum_{u=0}^{l-r} \mathrm{S}^{(2)}(l-u-f,r-f)
       \cdot \sum_{v=0}^{u} {f+v-1\choose f-1} \cdot
       \mathrm{S}^{(1)}(u,v)
\end{equation}
with the initial conditions
\begin{equation}\label{ini_cond}
S^{(1)}(l,0)=\alpha\cdot \delta_{l,0}, \ S^{(2)}(l,0)=(1-\alpha)\cdot\delta_{l,0}.
\end{equation}
\end{theorem}
The following denotations are used:
$$
{\cal{M}}^{(N,p,\alpha)}_k\!=\! \int \lambda^k d \sigma\left({\lambda;
A^{(N,p,\alpha)}}\right), \ M^{(N,p,\alpha)}_k=\mathbb{E}{\cal{M}}^{(N,p,\alpha)}_k,
$$
$$
C^{(N,p,\alpha)}_{k,m}= \mathbb{E}\left\{{\cal{M}}^{(N,p,\alpha)}_{k}\cdot {\cal{M}}^{(N,p,\alpha)}_{m} \right\}-\mathbb{E}\left\{{\cal{M}}^{(N,p,\alpha)}_{k} \right\}\cdot \mathbb{E}\left\{{\cal{M}}^{(N,p,\alpha)}_{m} \right\} .
$$

 Theorem \ref{main_thm} is a corollary of Theorem \ref{thm:1}
\begin{theorem} \label{thm:1}
Assuming  conditions (2.4),

(i) Correlators are vanished in the limit
\begin{equation}
C^{(N,p,\alpha)}_{k,m}\le \fracd{C(k,m,p,\alpha)}{N} , \ N\to \infty \ \forall\  k,m \in \mathbb{N}.
\end{equation}

(ii) The limit of s-th moment exists for all $s\in \mathbb{N}$
\begin{equation}\label{5}
  \lim_{N \to \infty}M^{(N,p, \alpha)}_s =  \left\{ \begin{array}{ll}
\!\sum^{k}_{i=0}\left( S^{(1)}(k,i)+ S^{(2)}(k,i)\right),&
\textrm{if} \ s=2k ,
\\0,& \textrm{if } \ s=2k-1 ,\\ \end{array} \right. ,
\end{equation}
where  numbers
  $S^{(1)}(k,i)$,  $S^{(2)}(k,i)$
are determined by the system of recurrent relations (\ref{ms_r}) -(\ref{ms_r2})
with the initial conditions  (\ref{ini_cond}).

(iii) The limiting moments $\left\{m^{(p,\alpha)}_{k}\right\}_{k=1}^{\infty}$ obey Carleman's condition
\begin{equation}
\sumd_{k=1}^{\infty}\fracd{1}{\sqrt[2k]{m^{(p,\alpha)}_{k}}}=\infty
\end{equation}
\end{theorem}

\section{Proof of Theorem 1}
\subsection{Walks and contributions}

\quad\ Using independence of families $\Xi$ and
$\Lambda^{(p)}_N$, we have {\setlength\arraycolsep{1pt}
\begin{eqnarray}
  M^{(N,p)}_k\!&=&\!\int {\bf E} \{ \lambda^k d \sigma_{A^{(N,p,\alpha)}}
\} \!=\!{\bf E} \left(\frac{1}{N}\sum_{i=1}^N
[\lambda^{(N,p,\alpha)}_i]^k \right)\!=\! \frac{1}{N} {\bf E}\left(Tr
[A^{(N,p,\alpha)}]^k\right) \!=\! \nonumber \\ & =&\! \frac{1}{N}
\sum^{N}_{j_1=1} \sum^{N}_{j_2=1} \ldots
  \sum^{N}_{j_{k}=1} {\bf E} \left( A^{(N,p,\alpha)}_{j_1,j_2}
  A^{(N,p,\alpha)}_{j_2,j_3} \ldots A^{(N,p,\alpha)}_{j_{k},j_1}
  \right) \!=\! \nonumber \\
\label{base} &=&\! \frac{1}{N} \sum^{N}_{j_1=1} \sum^{N}_{j_2=1}
\ldots
  \sum^{N}_{j_{k}=1} {\bf E} \left( a_{j_1,j_2}
  a_{j_2,j_3} \ldots a_{j_{k},j_1} \right) \cdot \nonumber \\ & &
  \cdot {\bf E} \left( d^{(N,p)}_{j_1,j_2}
  d^{(N,p)}_{j_2,j_3} \ldots d^{(N,p)}_{j_{k},j_1}
  \right) \cdot  \xi^{(N,\alpha)}_{j_1,j_2} \cdot
  \xi^{(N,\alpha)}_{j_2,j_3} \cdot \ldots \cdot\xi^{(N,\alpha)}_{j_{k},j_1},
 \end{eqnarray}
 where
 $$
\xi^{(N,\alpha)}_{ij}\!=\!\left\{ \begin{array}{ll} 1,&
\textrm{if} \  (i \in I_{\alpha,N} \wedge j\notin I_{\alpha,N} ) \vee (i \notin I_{\alpha,N} \wedge j\in I_{\alpha,N} )  ,
\\0,& \textrm{otherwise} \\ \end{array}
\right.
$$

Consider $W^{(N)}_{k}$ the set of closed walks of $k$ steps over
the set $\overline{1,N}$:
$$
W^{(N)}_{k}\!=\!\{w\!=\!(w_1,w_2,\cdots,w_k,w_{k+1}=w_1):
\forall i \!\in\! \overline{1,k+1} \;\: w_i\!\in\!
\overline{1,N}\}.
$$
For $w\!\in\!W^{(N)}_k$ let us denote
  $a(w)\!=\!\prod_{i=1}^{k} a_{w_i,w_{i+1}}$,
 $d^{(N,p)}(w)\!=\!\prod_{i=1}^{k} d^{(N,p)}_{w_i,w_{i+1}}$ and $\xi^{(N,\alpha)}(w)\!=\!\prod_{i=1}^{k} \xi^{(N,\alpha)}_{w_i,w_{i+1}}$.
Then we have
\begin{equation}\label{m_ms1}
M^{(N,p)}_k\!=\!\frac{1}{N} \sum_{w\in W^{(N)}_k} {\bf E} a(w)
\cdot {\bf E} d^{(N,p)}(w)\cdot \xi^{(N,\alpha)}(w).
\end{equation}
Let $w\!\in\! W^{(N)}_k$
  and
$f,g \!\in\! \overline{1,N}\ $. Denote by $n_w(f,g)$ the number
of steps $f\to g$ and $g \to f$;
$$
n_w(f,g)=\#\{i \!\in\! \overline{1,k}:\; (w_i\!=\!f \ \wedge \
w_{i+1}\! =\!g)\vee (w_i\!=\!g \ \wedge \ w_{i+1}\!=\!f)\}.
$$
Then
$$
{\bf E}a(w)\!=\! \prod_{f=1}^{N} \prod_{g=f}^{N} X_{n_w(f,g)}.
$$

  Given $w\!\in\! W^{(N)}_k$,
let us define the sets $V_w=\cup_{i=1}^{k}\{w_i\}$ and
$E_w=\cup_{i=1}^{k}\{(w_i,w_{i+1})\},$ where $(w_i,w_{i+1})$ is a
non-ordered pair. It is easy to see that $G_w\!=\!(V_w,E_w)$ is a
simple non-oriented graph and the walk $w$ covers the graph
$G_w$. Let us call $G_w$ the skeleton of walk $w$. We denote by
$n_w(e)$ the number of passages of the edge $e$ by the walk $w$ in
direct and inverse directions. For
$(w_j,w_{j+1})\!=\!e_j\!\in\!E_w$ let us denote
$a_{e_j}\!=\!a_{w_j,w_{j+1}}\!=\!a_{w_{j+1},w_j}$. Then we obtain
$$
  {\bf
E}a(w)\!=\!\prod_{e\in E_w} {\bf E}a^{n_w(e)}_e\!=\! \prod_{e\in
E_w} X_{n_w(e)}.
  $$
Similarly we can write
$$
  {\bf E}d^{(N,p)}(w)\!=\!\prod_{e\in E_w} {\bf
E}\left([d^{(N,p)}_e]^{n_w(e)} \right)\!=\! \prod_{e\in E_w}
\frac{p}{N}.
$$
Then, we can rewrite (\ref{m_ms1}) in the form
$$
M^{(N,p)}_k\!=\!\frac{1}{N}\sum_{w\in W^{(N)}_k} \xi^{(N,\alpha)}(w) \cdot \prod_{e\in E_w}
\frac{p\cdot X_{n_w(e)}}{N}\!=
$$
\begin{equation}\label{m_ms2}
=\!\sum_{w\in W^{(N)}_k} \xi^{(N,\alpha)}(w) \cdot\left(\frac{p^{|E_w|}}{N^{|E_w|+1}
}\prod_{e\in E_w}X_{n_w(e)} \right)\!=\!\sum_{w\in
W^{(N)}_k}\theta(w),
\end{equation}
where $\theta(w)$ is the contribution of the walk $w$ to the
mathematical
  expectation of the corresponding moment. To perform the limiting
transition
   $N\to\infty$ it is natural to separate $W^{(N)}_k$ into classes of
equivalence.
   Walks $w^{(1)}$ and $w^{(2)}$ are equivalent
  $ w^{(1)}\sim w^{(2)},\;$
if and only if there exists a bijection $\phi$ between the sets of
vertices $V_{w^{(1)}}$ and  $V_{w^{(2)}}$ such that for
$i=1,2,\ldots,k
  \;\; w^{(2)}_i\!\!=\!\!\phi(w^{(1)}_i)$
$$ w^{(1)}\sim w^{(2)}\;\Longleftrightarrow
\; \exists \phi: \ V_{w^{(1)}}\stackrel{bij}{\to}
  V_{w^{(2)}}:
  \;\forall \;i \!\in\! \overline{1,k+1}\;
\; w^{(2)}_i\!\!=\!\!\phi(w^{(1)}_i) \wedge {\mathbf{1}}_{I_{\alpha,N}}(w^{(2)}_i)={\mathbf{1}}_{I_{\alpha,N}}(\phi(w^{(1)}_i))$$
  Last condition requires that every vertex and it image be in same component. It's essential for further computations because  contribution of walk equals zero in the case when origin and end of some step are in the same component. Let us denote by $[w]$ the class of equivalence of walk $w$ and by
$C^{(N)}_k$ the set of such classes. It is obvious that if two
walks $w^{(1)}$ and $w^{(2)}$ are equivalent then their
contributions are equal.
$$
  w^{(1)}\sim w^{(2)}\;\Longrightarrow
\theta(w^{(1)})\!=\!\theta(w^{(2)})
$$
Cardinality of the class of equivalence $[w]$ is equal the number
of all mappings $\phi:V_w \to  \overline{1,N}$ such that $\phi(V_{1,w}) \subset  I_{\alpha,N}$ and $\phi(V_{2,w}) \subset  \overline{1, N}\setminus I_{\alpha,N}$ (, where $V_{1,w}=V_{w}\cap I_{\alpha,N}$ and  $V_{2,w}=V_{w}\setminus I_{\alpha,N}$)  i.e. $[\alpha\cdot N]
\cdot ([\alpha\cdot N]-1) \cdot \ldots \cdot ([\alpha\cdot N]-|V_{1,w}|+1)\cdot (N-[\alpha\cdot N])
\cdot (N-[\alpha\cdot N]-1) \cdot \ldots \cdot (N-[\alpha\cdot N]-|V_{2,w}|+1)$. Then we can rewrite
(\ref{m_ms2}) in the form
$$M^{(N,p)}_k\!=\!\sum_{w\in W^{(N)}_k}\xi^{(N,\alpha)}(w)
\left(\frac{p^{|E_w|}}{N^{|E_w|+1} }\prod_{e\in
E_w}X_{n_w(e)} \right)\!=$$
$$
=\!\sum_{[w]\in CW^{(N)}_k} \xi^{(N,\alpha)}(w)\prod_{e\in E_w}X_{n_w(e)} \left(\frac{[\alpha\cdot N]
\cdot ([\alpha\cdot N]-1) \cdot \ldots\cdot ([\alpha\cdot N]-|V_{1,w}|+1)} {N^{|E_w|+1}\cdot
p^{-|E_w|}}\cdot \right.
$$
\begin{equation}\label{m_ms3}
\left. (N-[\alpha\cdot N])
\cdot (N-[\alpha\cdot N]-1) \cdot \ldots \cdot (N-[\alpha\cdot N]-|V_{2,w}|+1)\right)\!=\!\sum_{[w]\in
CW^{(N)}_k} \hat{\theta}([w]).
\end{equation}
In second line of (\ref{m_ms3}) for every class $[w]$ we choose arbitrary walk $w$ corresponding to this class of equivalence.
\subsection{Minimal and essential walks}

   Class of walks $[w]$ of $CW^{(N)}_k$ has at most k vertices.
    Hence, $CW^{(1)}_k
\subset CW^{(2)}_k \subset \ldots \subset CW^{(i)}_k \subset
\ldots CW^{(k\cdot \min(\alpha,1-\alpha)^{-1})}_k = CW^{(k\cdot \min(\alpha,1-\alpha)^{-1}+1)}_k= \ldots$. It is natural to denote
  $CW_k=CW^{(\min(\alpha,1-\alpha)^{-1})}_k$. Then (\ref{m_ms3}) can be written as

\begin{equation}\label{m_ms5}
m^{(p)}_k \!=\!\lim_{N \to \infty} \sum_{[w]\in CW_k} \xi^{(N,\alpha)}(w)\cdot \alpha^{|V_{1,w}|}\cdot (1-\alpha)^{|V_w|-|V_{1,w}|}
\left(N^{|V_w|-|E_w|-1}\prod_{e\in
E_w}\fracd{ X_{n_w(e)}}{p^{-1}}\right).
\end{equation}
The set $CW_k$ is finite. Regarding this and (\ref{m_ms5}), we
conclude that the class $[w]$ has non-vanishing contribution,
if $|V_w|-|E_w|-1 \!\geq \! 0$ and $w$ is a bipartite walk  through the complete bipartite graph $K_{I_{\alpha,N},\overline{1,N}\setminus I_{\alpha,N}}$. But for each simple connected
graph $G=(V,E)$  $|V_w|\! \leq \! |E_w|+1$, and the equality takes
place if and only if the graph $G$ is
  a tree.

It is convenient to use a notion of minimal walk.
\begin{definition}
  The walk $w$ is a minimal walk, if $w_1$ (the root of walk) has
  the number 1 and the number of each new vertex is equal to the number
  of all already passed vertices plus 1.
\end{definition}
Let us denote the set of all minimal walks of $W^{(N)}_{k}$ by
$MW^{(N)}_{k}$.
\vskip 0.5cm

\noindent {\bf Example 1.} The sequences  (1,2,1,2,3,1,4,2,1,4,3,1) and
(1,2,3,2,4,2,3,2,1,2,4,1,5,1) represent the   minimal walks.

\vskip 0.5cm
\begin{definition}
The  minimal walk $w$ that has a tree as
a skeleton is an essential walk.
\end{definition}
Let us denote the set of all essential  walks of $W^{(N)}_{k}$ by
$EW^{(N)}_{k}$.
 Therefore we can rewrite

(\ref{m_ms5}) in the form
\begin{equation}\label{m_ms6}
m^{(p)}_k \!=\! \sum_{w\in EW_k} \left(\theta_1(w)+\theta_2(w) \right).
\end{equation}
where
$$
\theta_1(w)=\alpha^{\beta(w)}\cdot (1-\alpha)^{|V_w|-\beta(w)}
\left(\prod_{e\in
E_w}\left(p\cdot  X_{n_w(e)}\right)\right),
$$
$$
\theta_2(w)=(1-\alpha)^{\beta(w)}\cdot \alpha^{|V_w|-\beta(w)}
\left(\prod_{e\in
E_w}\left(p\cdot  X_{n_w(e)}\right)\right),
$$
where  $\beta(w)$ is a number of such vertices $v$ that the distance between $v$ and the first vertex $w_1$ are even.
 The number of passages of each edge $e$
  belonging to the essential walk $w$ is even. Hence, the limiting
  mathematical expectation $m^{(p)}_k$ depends only on the even
  moments of random variable of $a$. It is clear that the limiting
mathematical
  expectation $\limd_{N \to \infty}M^{(N,p,\alpha)}_{2s+1}$ is equal to zero.

\subsection{First edge decomposition of essential walks}
Let us start with necessary definitions. The first vertex $w_1=1$
of the essential walk $w$ is called the root of the walk. We
denote it by $\rho$. Let us denote the second vertex $w_2=2$ of
the essential walk  $w$ by $\nu$. We denote by $l$ the half of
walk's length and by $r$ the number of steps of $w$ starting from
root $\rho$.
  In this subsection  we derive the recurrent
  relations by splitting of the walk (or of the tree) into two
  parts. To describe this procedure, it is convenient to consider
    the set of the essential walks of length $2l$ such that they
have $r$ steps starting from the root $\rho$.  We denote this set
by $\Lambda(l,r)$. One can see that this description is exact, in
the sense that it is  minimal and gives complete description of
the walks we need. Denote by $S^{(1)}(l,r)$, $S^{(2)}(l,r)$  the sum of contributions of
the walk of $\Lambda(l,r)$ with weights $\theta_1$ and $\theta_2$ respectively. Let us remove the edge
$(\rho,\nu)=(1,2)$
  from $G_w$ and denote by $\hat{G}_w$ the graph
obtained . The graph $\hat{G}_w$ has two components. Denote the
component that contains the vertex $\nu$ by $G_2$ and the
component containing the root $\rho$ by $G_1$. Add the edge
$(\rho,\nu)$ to the edge set of the tree $G_2$. Denote the result
of this operation by $\hat{G}_2$.
  Denote by $u$ the half of the walk's length
over the tree $G_2$ and by $f$ the number of steps $(\rho,\nu)$
in the walk $w$. It is clear that the following inequalities hold
for all essential walks (excepting the walk of length zero) $1\leq
f\leq r$, $r+u\leq l$. Let us denote by $\Lambda_1(l,r,u,f)$ the
set of the essential walks with fixed parameters $l$, $r$, $u$,
$f$ and  by $S^{(1)}_1(l,r,u,f)$ ($S^{(2)}_1(l,r,u,f)$) the sum of contributions of the walks
of $\Lambda_1(l,r,u,f)$ with weight $\theta_1$ ($\theta_2$). Denote by $\Lambda_2(l,r)$ the set of
the essential walks of $\Lambda(l,r)$ such that their skeleton
has only one edge attached the root $\rho$.
  Also we denote by $S^{(1)}_2(l,r)$ and $S^{(2)}_2(l,r)$ the sum of weights  $\theta_1$ and $\theta_2$  respectively of the
walk of $\Lambda_2(l,r)$. Now we can formulate the first lemma of
decomposition. It allows express $S^{(1)}$, $S^{(2)}$ as  functions of the $S^{(1)}$, $S^{(2)}$, $S_2^{(1)}$, $S_2^{(2)}$.

\begin{lemma}[First decomposition lemma] The following relation holds
\begin{equation}\label{l11}
\mathrm{S}^{(1)}(l,r)\!=\!\sum_{f=1}^{r}\sum_{u=0}^{l-r}\mathrm{S}^{(1)}_1
(l,r,u,f),
\end{equation}
\begin{equation}\label{l11a}
\mathrm{S}^{(2)}(l,r)\!=\!\sum_{f=1}^{r}\sum_{u=0}^{l-r}\mathrm{S}^{(2)}_1
(l,r,u,f),
\end{equation}
where
\begin{equation}\label{l12}
\mathrm{S}^{(1)}_1(l,r,u,f) \!=\!\alpha^{-1}\cdot {r-1\choose
f-1}\cdot\mathrm{S}^{(1)}_{2}(f+u,f) \cdot
  \mathrm{S}^{(1)}(l-u-f,r-f),
\end{equation}
\begin{equation}\label{l12a}
\mathrm{S}^{(2)}_1(l,r,u,f) \!=\!(1-\alpha)^{-1}\cdot {r-1\choose
f-1}\cdot\mathrm{S}^{(2)}_{2}(f+u,f) \cdot
  \mathrm{S}^{(2)}(l-u-f,r-f).
\end{equation}
\end{lemma}
\

{\it Proof.} The first two equalities are obvious. The last two equalities follow
from
  the bijection
$$\mathrm{\Lambda}_1(l,r,u,f) \stackrel{bij}{\to}
\mathrm{\Lambda}_{2}(f+u,f) \times
  \mathrm{\Lambda}(l-u-f,r-f) \times $$
\begin{equation}\label{bij}
    \times \mathrm{\Theta}_{1}(r,f) ,
\end{equation}
where $\mathrm{\Theta}_1(r,f)$ is the set of sequences of 0 and 1
of
  length $r$ such that there are exactly $f$ symbols 1 in the
  sequence and the first symbol is 1.

     Let us construct this
  mapping $F$. Regarding one particular essential walk $w$ of
  $\Lambda_1(l,r,u,f)$, we consider the first edge $e_1$ of the
  graph $G_w$ and separate $w$ in two parts, the left and the
  right ones with respect to this edge $e_1$. Then we add a special
  code that determines the transitions from the left part to the
  right one and back through the root $\rho$.
  Obviously these two parts are walks, but not necessary minimal
  walks. Then  we minimize these walks. This decomposition is
  constructed by the following algorithm. We run over $w$ and
  simultaneously draw the left part, the right part, and code. If
  the current step belongs to $G_l$, we add it to the first part,
  otherwise we add this step to the second part. The code is
  constructed as follows. Each time the walk leaves the root the
  sequence is enlarged by one symbol. If
  current step is $\rho \to \nu$ and "0" otherwise,  this symbol is "1".
   It is clear that
  the first element of the sequence is "1", the number of signs "1" is equal to
   $f$, and the  full length of the sequence is $r$. Now we
minimize the left and the right parts. Thus, we have constructed
the decomposition of the essential walk $w$ and the mapping $F$. The weight $\theta_1(w)$($\theta_2(w)$) of the
essential walk is  multiplicative with respect to edges and vertices. In factors $\mathrm{S}^{(1)}_{2}(f+u,f)$,
  $\mathrm{S}^{(1)}(l-u-f,r-f)$ we twice count multiplier corresponding to the root, so we need add factor $\alpha^{-1}$ in (\ref{l12}).

\vskip 1cm

\noindent {\bf Example 2.} For
$w=(1,2,1,2,3,2,1,4,1,2,5,2,1,4,6,4,1,2,5,2,3,2,3,2,1,4,1)$ the
left part, the right one, and the code are
$(1,2,1,2,3,2,1,2,4,2,1,2,4,2,3,2,3,2,1)$,$\;(1,2,1,2,3,2,1,2,1)$,
$(1,1,0,1,0,1,0)$, respectively.

\vskip 1cm

Let us denote the left part by $( w^{(f)} )$ and the right part by
$( w^{(s)} )$. These parts are really walks with the root $\rho$.
For each edge $e$ in the tree $\hat{G}_2$ the number of passages
of $e$ of the essential walk $w$ is equal to the corresponding
number of passages of $e$ of the left part $( w^{(f)} )$. Also
for each edge $e$ belonging to the tree
  $G_1$ the number of passages of $e$
of essential walk $w$ is equal to the corresponding number of
passages of $e$ of the right part $( w^{(s)} )$. The weight of the
essential walk is  multiplicative with respect to edges. Then the
weight of the essential walk $w$ is equal to the product of
weights of left and right parts. The walk of zero length has unit
weight. Combining this with (\ref{bij}), we obtain
\begin{equation}\label{aux}
\mathrm{S}^{(1)}_1(l,r,u,f)= \alpha^{-1}\cdot\left| \mathrm{{\Theta}}_{1}(r,f) \right|
\cdot \mathrm{S}^{(1)}_2(f+u,f) \cdot \mathrm{S}^{(1)}(l-u-f,r-f),
\end{equation}
\begin{equation}\label{aux1}
\mathrm{S}^{(2)}_1(l,r,u,f)= (1-\alpha)^{-1}\cdot\left| \mathrm{{\Theta}}_{1}(r,f) \right|
\cdot \mathrm{S}^{(2)}_2(f+u,f) \cdot \mathrm{S}^{(2)}(l-u-f,r-f).
\end{equation}

Taking into account that $|\mathrm{\Theta}_{1}(r,f)|={r-1\choose
f-1}$, we derive from (\ref{aux})-(\ref{aux1})  (\ref{l12})-(\ref{l12a}).

Now let us prove that for any given  elements $w^{(f)}$ of
$\mathrm{\Lambda}_{2}(f+u,f)$, $w^{(s)}$ of
  $\mathrm{\Lambda}(l-u-f,r-f)$, and the sequence $\theta \in
  \mathrm{\Theta}_{1}(r,f)$, one can construct one and only one
  element $w$ of $\mathrm{\Lambda}_1(l,r,u,f)$. We do this
  with the following gathering algorithm. We go along either
$w^{(f)}$ or $w^{(s)}$ and simultaneously draw the walk $w$. The
switch from $w^{(f)}$ to $w^{(s)}$ and back is governed by the
code sequence $\theta$. In fact, this procedure is inverse to the
decomposition procedure described above up to the fact that
$w^{(s)}$ is minimal. This difficulty can be easily resolved for
example by coloring vertices of $ w^{(f)} $ and $ w^{(s)} $ in
red and blue colors respectively. Certainly, the common root of $
w^{(f)} $ and $ w^{(s)} $has only one color. To illustrate
the gathering procedures we give the following example.

\vskip 1cm

\noindent {\bf Example 3.} For $ w^{(f)}
=(1,2,1,2,3,2,1,2,4,2,1,2,4,2,3,2,3,2,1)$,$\; w^{(s)}
=(1,2,1,2,3,2,1,2,1), $ $\;\theta=(1,1,0,1,0,1,0)$ the gathering
procedure gives $\;w=(1,2,1,2,3,2,1,4,1,2,5,2,1,4,6,$
$4,1,2,5,2,3,2,3,2,1,4,1)$.

\vskip 1cm

It is clear that the decomposition and gathering are  injective
mappings. Their domains are finite sets, and therefore the
corresponding mapping (\ref{bij}) is bijective. This completes the
proof of Lemma 1.$\blacksquare$

\vskip 1cm

To formulate Lemma 2, let us give necessary definitions. We denote by $v$
the number of steps
  starting from the vertex $\nu$ excepting the steps
$\nu \to \rho$ and by $\Lambda_3(u+f,f,v)$ the
set of essential walks
   of $\Lambda_2(u+f,f)$ with fixed parameter $v$. Also we denote
   by $S^{(1)}_3(u+f,f,v)$ ($S^{(2)}_3(u+f,f,v)$) the sum of weights $\theta_1$ ($\theta_2$) of walks of
   $\Lambda_3(u+f,f,v)$. Let us denote by $G_{1,2}$ the
graph consisting of only one edge $(\rho,\nu)$ and by
$\Lambda_4(f)$ the set of essential walks of length $2f$ such that
their skeleton coincides with the graph $G_{1,2}$. It is clear that
$\Lambda_4(f)$ consists of the only one walk (1,2,1,2,\ldots,2,1)
of weight $\frac{X_{2f}}{p^{-1}}$. The previous lemma allows us to
express $\mathrm{S}^{(1)}_2$, $\mathrm{S}^{(2)}_2$ as  functions of $\mathrm{S}^{(1)}$, $\mathrm{S}^{(1)}$. The next
lemma allows to express $\mathrm{S}^{(1)}_2$, $\mathrm{S}^{(2)}_2$ as  functions of
$\mathrm{S}^{(1)}$, $\mathrm{S}^{(2)}$. Thus, two lemmas allow us to express $\mathrm{S}^{(1)}$, $\mathrm{S}^{(2)}$
as  functions of $\mathrm{S}^{(1)}$, $\mathrm{S}^{(2)}$.
\begin{lemma}[Second decomposition lemma]
\begin{equation}\label{l21}
\mathrm{S}^{(1)}_2(f+u,f)=\sum_{v=0}^{u} \mathrm{S}^{(1)}_3(f+u,f,v)
\end{equation}
\begin{equation}\label{l21a}
\mathrm{S}^{(2)}_2(f+u,f)=\sum_{v=0}^{u} \mathrm{S}^{(2)}_3(f+u,f,v)
\end{equation}
\begin{equation}\label{l22}
\mathrm{S}^{(1)}_3(f+u,f,v)\!=\! \alpha\cdot{f+v-1\choose f-1}\cdot
\frac{X_{2f}}{p^{-1}} \cdot \mathrm{S}^{(2)}(u,v)
\end{equation}
\begin{equation}\label{l22a}
\mathrm{S}^{(2)}_3(f+u,f,v)\!=\! (1-\alpha)\cdot{f+v-1\choose f-1}\cdot
\frac{X_{2f}}{p^{-1}} \cdot \mathrm{S}^{(1)}(u,v)
\end{equation}
\end{lemma}
The first two equalities are trivial, the second two follow from the
bijection
\begin{equation}\label{bij2}
\mathrm{\Lambda}_3(f+u,f,v) \stackrel{bij}{\to}
\mathrm{\Lambda}(u,v)
  \times \mathrm{\Lambda}_4(f) \times \mathrm{ \Theta}_{2}(f+v,f),
\end{equation}
where
  $\mathrm{\Theta}_{2}(f+v,f)$ is the set of sequences of $0$ and $1$
of
  length $f+v$ such that there are exactly $f$ symbols 1 in the
  sequence and last symbol of it is 1. The proof is analogous to
  the proof of the first decomposition lemma. The factor $\alpha$ in (\ref{l22}) is a contribution of the root in the weight.


Combining these two decomposition lemmas and changing the order of
summation, we get the recurrent relations (\ref{ms_r})-(\ref{ms_r2})
with the initial conditions (\ref{ini_cond}) .




\begin{thebibliography}{99}








\bibitem{BG1} M.Bauer and O.Golinelli. Random incidence matrices:
spectral density at zero energy, Saclay preprint T00/087;
cond-mat/0006472



\bibitem{B} B. Bollobas {\it Random Graphs }   Acad. Press (1985)


\bibitem{BG2} M.Bauer and O.Golinelli. Random incidedence matrices:
moments and
spectral density, J.Stat. Phys. {\bf 103}, 301-336, 2001



\bibitem{Ch} Fan R.K. Chung, {\it Spectral Graph Theory}  {\bf }
AMS (1997)


\bibitem{CDS}  D.M. Cvetkovi$ \acute{c}$, M.Doob, and H.Sachs.
        \textit{Spectra of Graphs}, Acad. Press (1980)



\bibitem{E1} S.N. Evangelou. Quantum percolation and the Anderson
transition in dilute systems, {\it Phys. Rev. B }  {\bf 27} (1983)
1397-1400


\bibitem{E2} S.N. Evangelou and E.N. Economou. Spectral density
singularities, level statistics, and localization in
sparse random matrices,
  {\it Phys. Rev. Lett. }  {\bf 68} (1992) 361-364

\bibitem{E3} S.N. Evangelou. A numerical study of sparse
random matrices,  {\it J. Stat. Phys. }  {\bf   69}
(1992) 361-383

\bibitem{FM:96} Y.V.Fyodorov, A.D.Mirlin. Strong eigenfunction
correlations near the Anderson localization transition.
arXiv:cond-mat/9612218 v1




\bibitem{GR} Ch. Godzil, G. Royle, {\it  Algebraic Graph Theory.}  {\bf }
Springer-Verlag, New York (2001)


\bibitem{JLR}  S. Janson, T. \L uczak, A. Rucinski, {\it Random Graphs. }
{\bf }  John Wiley \& Sons, Inc. New York (2000)

\bibitem{JMRR} D. Jacobson, S.D. Miller, I. Rivin,
and Z. Rudnick. Eigenvalue spacing for regular graphs,
in: {\it Emerging applications of number theory. } Ed. D.A. Hejhal et al.
{\bf } Springer-Verlag  (1999)








\bibitem{KSV} Khorunzhy O., Shcherbina M., and Vengerovsky V. Eigenvalue distribution of large weighted random graphs, J. Math. Phys. {\bf 45}  N.4: (2004), 1648-1672.





\bibitem{MF:91} A.D.Mirlin, Y.V.Fyodorov. Universality of the
level correlation function of sparce random matrices,
  J.Phys.A:Math.Jen.{\bf 24}, (1991), 2273-2286.



\bibitem{RB:88} G.J. Rodgers  and A.J. Bray. Density of states of a
sparse random matrix, Phys.Rev.B {\bf 37}, (1988), 3557-3562.

\bibitem{RD:90}  G.J. Rodgers and C. De Dominicis. Density of states of
sparse random matrices,
J.Phys.A:Math.Jen.{\bf 23}, (1990), 1567-1566.


\bibitem{W} E.P.Wigner. On the distribution of the roots of
certain symmetric matrices, Ann.Math. {\bf 67}: (1958), 325-327.



\end{thebibliography}
\end{document}